\documentclass{emulateapj}
\usepackage{amsmath}
\usepackage{epsfig}
\usepackage{natbib}
\usepackage{color}

\newcommand{\msun}{\ensuremath{\mathrm{\,M}_{\odot}}}

\newcommand{\kpc}{\ensuremath{\mathrm{\,kpc}}}

\newcommand{\beq}{\begin{equation}}
\newcommand{\eeq}{\end{equation}}
\shortauthors{Walker et al.}

\begin{document}
\journalinfo{Accepted for publication in \textit{The Astrophysical Journal Letters}}
\title{Dark matter in the classical dwarf spheroidal galaxies:
a robust constraint on the astrophysical factor for $\gamma$-ray flux calculations}

\author{M.G. Walker$^{\dagger}$\altaffilmark{1,2}, C. Combet\altaffilmark{3},
J.A. Hinton\altaffilmark{3}, D. Maurin$^\ddagger$\altaffilmark{4,3,5}, M.I. Wilkinson\altaffilmark{3}}
\email{$^\dagger$mwalker@cfa.harvard.edu}
\email{$^\ddagger$dmaurin@lpsc.in2p3.fr}

\altaffiltext{1}{Institute of Astronomy, University of Cambridge, Madingley Road, Cambridge, CB3 0HA, UK}
\altaffiltext{2}{Harvard-Smithsonian Center for Astrophysics, 60 Garden St., Cambridge, MA 02138, USA}
\altaffiltext{3}{Dept. of Physics and Astronomy, University of Leicester, Leicester, LE1 7RH, UK}
\altaffiltext{4}{Laboratoire de Physique Subatomique et de Cosmologie,
CNRS/IN2P3/INPG/Universit\'e Joseph Fourier Grenoble 1,
53 avenue des Martyrs, 38026 Grenoble, France}
\altaffiltext{5}{Institut d'Astrophysique de Paris, CNRS/Universit\'e Pierre et Marie Curie, 98 bis bd Arago, 75014 Paris, France} 

\begin{abstract}

We present a new analysis of the relative detectability of dark matter annihilation in the Milky Way's eight `classical' dwarf spheroidal satellite galaxies.  Ours is similar to previous analyses in that we use Markov-Chain Monte Carlo techniques to fit dark matter halo parameters to empirical velocity dispersion profiles via the spherical Jeans equation, but more general in the sense that we do not adopt priors derived from cosmological simulations.  We show that even without strong constraints on the shapes of dSph dark matter density profiles (we require only that the inner profile satisfies $-\lim_{r\rightarrow 0}\mathrm{d}\ln\rho/\mathrm{d}\ln r \leq 1$), we obtain a robust and accurate constraint on the astrophysical component of a prospective dark matter annihilation signal, provided that the integration angle is approximately twice the projected half-light radius of the dSph divided by distance to the observer, $\alpha_{\rm int}\sim 2r_h/d$.  Using this integration angle, which represents a compromise between maximizing prospective flux and minimizing uncertainty in the dSph's dark matter distribution, we calculate the relative detectability of the classical dSphs by ground- and space-based $\gamma$-ray observatories.
\end{abstract}


\section{Introduction}

Given their large dynamical mass-to-light ratios ($M/L_V\ga 10^{1-3}[M/L_V]_{\odot}$, e.g., \citealt{aaronson83}, \citealt{mateo98} and references therein) and small baryonic contents ($L_V\sim 10^{3-7}L_{V,\odot}$), the Milky Way's dwarf spheroidal (dSph) satellite galaxies have become targets of interest in searches for high-energy photons that might be released in self-annihilation \citep[e.g.][]{1990Natur.346...39L,stoehr03,Evans:2003sc} or decay \citep[e.g.,][]{boyarsky06} events involving the as-yet unidentified particle(s) that constitute cosmological dark matter.  Such an `indirect' detection of dark matter has thus far remained elusive \citep[e.g., ][]{abdo10,acciari10,aharonian10,boyarsky10,seg1magic11}, but recently-commissioned and planned instruments will soon explore the sky with unprecedented sensitivity \citep{atwood09, cta:concept}.  

Any prospective signal depends on the nature of dark matter and its distribution in the emitting object.  Stellar kinematic data can provide information about the distribution of dark matter in dSphs that, combined with some future observation of annihilation/decay products, would then amount to information about the nature of dark matter.  High-quality stellar kinematic data sets now exist for each of the Milky Way's eight `classical' dSph satellites \citep[e.g.,][]{kleyna02,munoz06,koch07,sohn07,mateo08,walker09a}, and several groups have used these data to predict annihilation signals \citep[e.g.,][]{2009JCAP...01..016B,pieri09}).  Several published calculations \citep[e.g.,][]{strigari08b,martinez09,kuhlen10} of indirect detection signals from dSphs begin by adopting `cosmological priors', effectively assuming that dSphs occupy dark matter halos that have structural parameters (e.g. virial mass, scale radius and asymptotic behavior of the inner density profile) characteristic of halos produced in dissipationless N-body simulations of galaxy formation that assume dark matter is `cold' \citep[e.g.,][]{navarro97,navarro04,kuhlen08,springel08}.  

Given a lack of direct observational constraints on the density profiles of dSph dark matter halos, here we provide a more general analysis with the goal of examining the robustness and veracity with which a Jeans analysis of dSph stellar-kinematic data can constrain the astrophysical component of a dark matter annihilation signal.  We adopt minimal assumptions about the structural properties of dSph dark matter halos and we test our method on artificial data drawn from physical distribution functions.  In this context we identify the optimal integration angle that balances interests in maximizing flux and minimizing uncertainty in the astrophysical component of a prospective annihilation signal.  Finally, we compare the classical dSphs in terms of the relative detectability (absolute detectability depends on unknown particle physics) of their dark matter annihilation signatures. In subsequent contributions we revisit this topic in greater detail, considering the impact of (sub-) substructure and providing a thorough discussion of the instrumental response (Charbonnier et al. in preparation). 

\section{\emph{J}-factor and confidence intervals }

The differential $\gamma$-ray flux (units cm$^{-2}$~s$^{-1}$~GeV$^{-1}$) received on Earth in solid angle $\Delta\Omega$ from dark matter annihilations is given by
\begin{equation}
     \frac{d\Phi_{\gamma}}{dE_{\gamma}}
        = \frac{1}{4\pi}\frac{\langle\sigma v\rangle}{2m_{\chi}^{2}}
          \frac{dN_{\gamma}}{dE_{\gamma}} \times J(\Delta\Omega),
\label{eq:flux}
\end{equation}
where $m_{\chi}$ is the dark matter particle mass, $\langle \sigma v\rangle$ is the velocity-averaged annihilation cross section, $dN_{\gamma}/dE_{\gamma}$ is the energy spectrum of annihilation products and 
\beq
J(\Delta\Omega)=\int_{\Delta\Omega}\int \rho^2 (l,\Omega) \,dld\Omega\;.
\label{eq:J}
\eeq 
This `$J$-factor' represents the astrophysical contribution to the signal and is specified by the integral of the squared dark matter density, $\rho^2(l,\Omega)$, over line of sight $l$ and solid angle $\Omega$.

In order to allow for a broad range of dark matter density profiles, we adopt the \citet{zhao96} generalization (i.e., the $\alpha,\beta,\gamma$ model) of the \citet{hernquist90} and \citet[`NFW' hereafter]{navarro96} profiles,
\begin{equation}
  \rho(r)=\rho_s\biggl (\frac{r}{r_s}\biggr )^{-\gamma}\biggl [1+\biggl (\frac{r}{r_s}\biggr )^{\alpha}\biggr ]^{\frac{\gamma-\beta}{\alpha}}.
  \label{eq:hernquist1}
\end{equation}
We do not consider subclumps in the dSphs, as these substructures are not expected to contribute significantly to the $J$-factor (\citealt{springel08}, Charbonnier et al., in preparation). 

\subsection{Toy-model behavior}
The $J$-factor depends primarily on the inner logarithmic slope $\gamma\equiv -\lim_{r\rightarrow 0}\mathrm{d}\ln\rho / \mathrm{d}\ln r$, scale radius $r_s$, and normalization $\rho_s$ of the dark matter halo.  Only two of these three quantities are independent for the classical dSphs, such that masses within 300~pc can be reasonably well-constrained at $M_{300}\approx10^7 M_\odot$ \citep{2008Natur.454.1096S}. Here we find that $J$-factors for the classical dSphs depend only weakly on the exact profile as long as the inner profile has slope $0\leq \gamma\leq1$ ($\gamma=0$ corresponds to a `core' of constant density, whereas $\gamma=1$ corresponds to an NFW `cusp'). This independence can be understood using the point-like approximation formula for $J$, expressed as a function of dSph distance $d$ and integration angle\footnote{To avoid confusion with halo parameter $\alpha$ appearing in Equation \ref{eq:hernquist1}, we use subscripts to denote integration angles, e.g., $\alpha_{\rm int}$ and $\alpha_c$.} $\alpha_{\rm int}$ (with $\Delta\Omega = 2\pi(1-\cos\alpha_{\rm int})$)
\beq J_{\rm   point-like}(\alpha_{\rm int})=\frac{4\pi}{d^2}\int_{0}^{(\alpha_{\rm
    int}d)} r^3 \rho^2(r) \,d\ln(r).
\label{eq:Japprox}
\eeq
\begin{figure}[!t]
\begin{center}
\includegraphics[angle=0,scale=0.4]{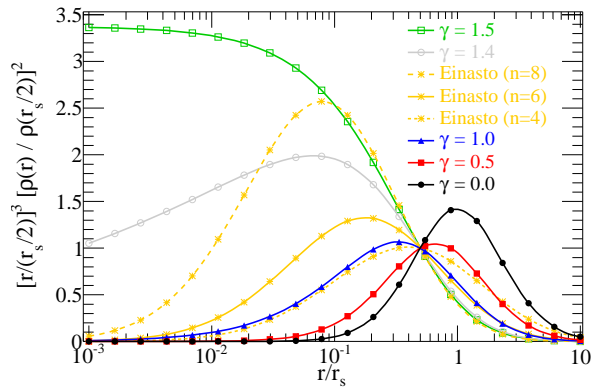}
\caption{Integrand appearing in Eq.~(\ref{eq:Japprox}) to calculate the $J$-factor for ($1,3,\gamma$) dark matter profiles.  For comparison we show the integrand calculated for the Einasto profile.  In this representation, $J_{\rm point-like}$ is directly proportional to the area under each curve.}
\label{fig:fig1}
\end{center}
\end{figure}
The shape of the integrand in Equation \ref{eq:Japprox} (using a logarithmic integration step) is plotted in Fig.~\ref{fig:fig1} for several $(\alpha,\beta,\gamma)=(1,3,\gamma)$ dark matter profiles (i.e., NFW with a free inner slope).  Each is normalized to the value at $r=r_s/2$, a radius that is expected to be of the same order of magnitude as the typical radii of stellar orbits.  For $r\ga r_s$, the contribution of the integrand quickly vanishes in all cases as the integrand drops at least as fast as $r^{-4}$.  For $\gamma=1.5$, the integral diverges in the inner parts unless one imposes a cutoff (such a cutoff is physically motivated by the need to balance rates of annihilation and gravitational infall, \citealt{1992PhLB..294..221B}).

Otherwise, all the integrands of profiles with $0<\gamma\leq1$ are peaked and have similar areas under their curves.  For $1<\gamma<1.4$, the integral is larger by a factor of a few, with the exact value depending on the choice of normalization.  Figure \ref{fig:fig1} also shows the integrand calculated for the `Einasto profile' (e.g., \citealt{einasto89}) that, with respect to NFW and by virtue of its variable inner slope, provides slightly better fits to simulated cold dark matter halos \citep{navarro04}.  The three orange curves in Figure \ref{fig:fig1} correspond to the Einasto profiles (indices $n=4,6,8$) used by \citet{navarro04} to fit the density profiles of low-mass cold dark matter halos.  The integrands from Equation \ref{eq:Japprox} corresponding to the plotted Einasto profiles (chosen here to have $-\mathrm{d}\ln\rho/\mathrm{d}\ln r=2$ at the scale radius of the plotted NFW profile) behave similarly in Figure \ref{fig:fig1} to the $\alpha,\beta,\gamma$ models plotted with $\gamma\leq 1.5$.  

\subsection{Jeans/MCMC analysis of classical dSphs}
Nearly all dSph stars with measured velocities lie within a few half-light radii\footnote{For consistency with \citet{walker09d} we define $r_{h}$ as the radius of the circle enclosing half of the dSph stellar light as seen in projection.  Elsewhere this radius is commonly referred to as the `effective radius'.} of the dSph center ($r_{h}\sim$ a few $\times 10^2$ pc for the Milky Way's classical dSphs).  Various published analyses have used the Jeans equation to show that the observed flatness of dSph velocity dispersion profiles \citep{walker07b} leads to a constraint on $M(r_{h})$---the mass enclosed within a sphere of radius $r_{h}$---that is insensitive to assumptions about either anisotropy or the structural parameters of the dark matter halo as specified, e.g., by $\alpha,\beta,\gamma$ \citep{walker09d,wolf10,amorisco10}.  For an indirect detection experiment, locating the \textit{optimal} integration angle, denoted $\alpha_c$, on a resolved target amounts to finding a compromise between maximization of signal (in the presence of background) and minimization of model-dependence in the corresponding astrophysical factor $J_{\alpha_c}$---and hence minimization of the uncertainty on the derived values of $m_{\chi}$ and $\langle \sigma v\rangle$.  For the Plummer models often projected to fit dSph surface brightness profiles, 80\% of the projected stellar mass lies within $2r_{h}$.  An integration angle $\alpha_{\rm int}=\alpha_c \sim  2r_{h}/d$ therefore represents a reasonable compromise between flux collection and robust estimation of the $J$-factor.
\begin{figure}[!t]
\begin{center}
\includegraphics[angle=0,scale=0.4]{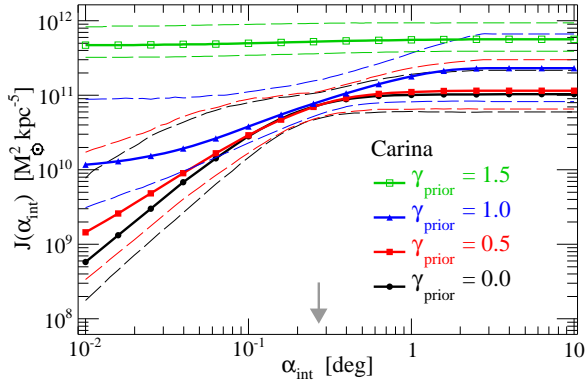}
\caption{$J$ median value and 95\% CLs from four independent MCMC
  analyses of Carina, holding the parameter $\gamma$ fixed at the
  values shown. The arrow identifies $\alpha_{\rm int}=\alpha_c \sim 2r_h/d$.}
\label{fig:fig2}
\end{center}
\end{figure}

We verify the robustness of Jeans estimates of $J_{\alpha_c}$---i.e., the $J$-factor corresponding to integration angle $\alpha_{\rm int}=\alpha_{c}\sim 2r_h/d$---by repeating the Markov-Chain Monte Carlo (MCMC) analysis of \citet[W09 hereafter]{walker09d} for the Milky Way's eight classical dSphs.  This analysis assumes virial equilibrium, spherical symmetry, constant velocity anisotropy, a dark matter density profile given by Eq.~(\ref{eq:hernquist1}), and a stellar luminosity profile given by a Plummer model, $L(r)/L_{\rm tot}=(r/r_{h})^3(1+[r/r_{h}]^2)^{-3/2}$, with half-light radius derived from the structural parameters originally published by \citet{ih95}.  We use the same velocity dispersion profiles displayed in Figure 1 of \cite{walker09d} (estimated under the assumption that the velocity distribution within each bin is Gaussian, thereby discarding dynamical information that might be contained in non-Gaussian distributions; e.g., \citealt{lokas05}), but our current analysis differs in the following ways.  First, we allow the outer logarithmic slope of the dark matter density profile to remain a free parameter, $\beta$, for which we take a flat prior over the range $3\leq \beta \leq 7$ (this change has no significant impact on our results).  Second, while we eventually allow $\gamma$ to vary freely between $0$ and $1$ (section \ref{sec:detection}), we perform several initial analyses of each dSph in which we fix the inner slope of the dark matter density profile at specific values of $\gamma=0$ (a constant density core), $\gamma=0.5$ (a mild cusp), $\gamma=1$ (NFW cusp) and $\gamma=1.5$ (a steep cusp).  Third, after the MCMC chains converge (see \citet{walker09d} for complete details of the MCMC procedure), we perform a numerical integration of Eq.~(\ref{eq:J}) that yields a value of the $J$-factor for each set of parameters present in the chains.  The distribution of these values then represents the observational constraint on the $J$-factor, given the modelling assumptions.

Figure \ref{fig:fig2} displays the median $J$-factor and 68\% confidence intervals (CIs) from our MCMC chains for the Carina dSph, where we have used integration angles $\alpha_{\mathrm{int}}$ between 0.01 and $5^\circ$ ($\alpha$, $\beta$, $r_s$, $\rho_s$ and the velocity anisotropy parameter free). In order to make explicit the effect of strong assumptions about the inner slope of the dark matter density profile, we show results from our four initial MCMC analyses with the slope held fixed at values $\gamma=0.0, 0.5, 1.0, 1.5$.  Small integration angles $\alpha_{\rm   int}\ll \alpha_c\sim 2r_h/d$ in Fig.~\ref{fig:fig2} correspond to $r\ll r_s$ in Fig.~\ref{fig:fig1}, so that the different values for the adopted $\gamma$ result in different median values and CIs for the $J$-factor.  However, for integration angles $\alpha_{\rm int}= \alpha_{c}\sim 2r_h/d$, the median $J$-factors and corresponding CIs converge to similar values regardless of whether we adopt $\gamma=0.0$, 0.5 or 1.0.  Adoption of a larger inner slope (e.g. $\gamma=1.5$ in green) gives a larger $J$-factor even as evaluated at $\alpha_{\rm int}= \alpha_{c}$, as expected based on the point-like approximation shown in Fig.~\ref{fig:fig1}. The other dSphs (not shown) exhibit the same behaviors, although with narrower/broader confidence intervals for dSphs in which more/less kinematic data are available.

\subsection{MCMC analysis of artificial data}
In order to examine its reliability, we apply our Jeans/MCMC analysis to artificial sets of stellar positions and velocities drawn from distribution functions of the form $L^{-2b} f(\varepsilon)$. Here $\varepsilon$ is energy, $L$ is angular momentum and $b$ is the velocity anisotropy parameter $b \equiv 1 - \sigma_{\rm t}^2/\sigma_{\rm r}^2$, where $\sigma_{\rm r}^2$ and $\sigma_{\rm t}^2$ are second moments of the velocity distribution in radial and (one of the) tangential directions, respectively. For $b\neq0$, these models give stellar distributions with constant velocity anisotropy~\citep[for a discussion see, e.g.,][]{bt08}. Following~\cite{cuddeford91}, once the halo model and stellar density are specified, an Abel inversion is used to determine the distribution function numerically. We generate a grid of halo models with $\gamma=0.1,0.5,1.0$; $r_{\rm h}/r_{\rm s}=0.1,0.5,1.0$ and $\beta=3.1$. For each model we consider isotropic ($b=0$) and anisotropic distributions with $b=0.25$ or $b=-0.75$.  We also generate a grid of models with $\alpha = 1.5, \beta=4.0$. All test halos are normalized to have mass $\sim 10^7$M$_\odot$ within 300pc. Stellar positions and velocities are generated directly from the full stellar distribution function using an accept-reject algorithm, and subsequently all velocities are scattered randomly to mimic observational errors of $\pm 2$ km s$^{-1}$. All artificial samples have 1000 stars, similar to the typical data set available for the classical dSphs.
\begin{figure}[!t]
\begin{center}
\includegraphics[angle=0,scale=0.4]{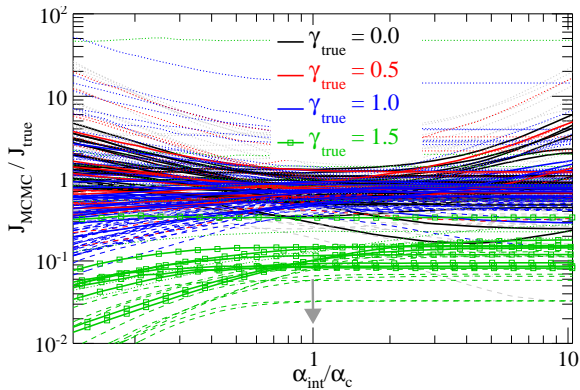}
\caption{From artificial data sets, ratio of the measured $J$-factor to the true $J$-factor, as a function of $\alpha_{\rm int}/\alpha_c$.  Lines are color-coded with respect to the true value of the slope of the inner density profile of the input model.  For these analyses, as for the real dSph data (section \ref{sec:detection}), we adopt a uniform prior over $0\leq \gamma\leq 1$ for the inner slope.  Solid lines indicate median values of the $J$-factors obtained in our MCMC chains, while dashed/dotted lines indicate lower/upper bounds on 95\% CIs.}
\label{fig:fig3}
\end{center}
\end{figure}

In our analysis of the artificial data sets, as in our final analysis of the real dSph data (section \ref{sec:detection}), we allow $\gamma$ to vary freely between $0\leq \gamma\leq 1$ (effectively adopting a flat prior over this range), reflecting the facts that 1) published kinematic analyses do not place strong constraints on this parameter (e.g. \citealt{koch07,walker09d}), and 2) current cosmological simulations suggest that $\gamma\la 1$ for the subhalos most readily associated with dSphs \citep[e.g.,][]{springel08}.  For 72 artificial dSphs, Fig.~\ref{fig:fig3} plots $J_{\rm MCMC}/J_{\rm true}$, the ratio of the measured (median and 95\% CIs) to the known input value of $J$, as a function of $\alpha_{\rm   int}/\alpha_c$.  For most of the test models, our estimates of $J_{\rm MCMC}(\alpha_c)$ lie within a factor of two of the correct values, even for models with extreme anisotropy.  However, for the most steeply cusped ($\gamma=1.5$) test models that violate our assumption that the inner slope satisfies $0\leq \gamma\leq 1$, we under-estimate the $J$-factor typically by a factor of $\ga 10$ (green curves in Figure \ref{fig:fig3}).  While the adoption of a less restrictive prior (e.g., allowing $0\leq \gamma\leq 2$) would avoid this potential bias, such an improvement would come at the cost of larger uncertainties for estimates of the $J$-factor in all halos.  Since current cosmological simulations \citep[e.g.,][]{springel08} and indirect arguments based on observations \citep[e.g.,][]{kleyna03,goerdt06,gilmore07} suggest that dSph dark matter halos have $\gamma\la 1$, we report results based on our more restrictive assumption that $0\leq \gamma\leq 1$.  Figure \ref{fig:fig3} illustrates that our estimates of the $J$-factor are as reliable as this assumption.  

\section{Detection prospects}
\label{sec:detection}

\begin{figure}[!t]
\begin{center}
\includegraphics[angle=0,scale=0.4]{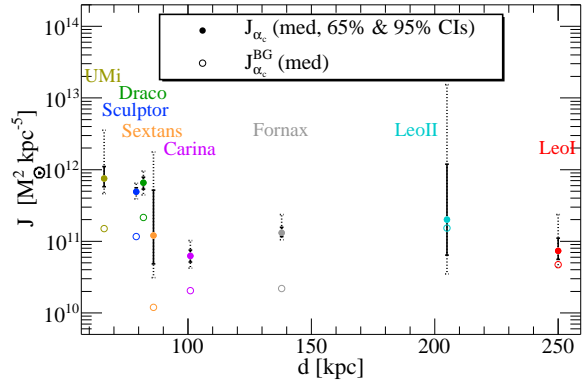}
\caption{Astrophysical factor as a function of the distance to the dSph. Filled circles, solid
and dashed error bars correspond respectively to the median value, 68\%, and 95\% CIs on $J_{\alpha_{c}}$
(where $\alpha_c$ is given in Table~\ref{tab:tab1}). Empty circles correspond to 
$J^{\rm BG}_{\alpha_{c}}=J_{\alpha_c}/\sqrt{1+(\alpha_c/0.1^{\circ})^2}$.}
\label{fig:fig4}
\end{center}
\end{figure}

Table \ref{tab:tab1} and Figure \ref{fig:fig4} indicate estimates of $J_{\alpha_c}$ we obtain for the real dSphs from our Jeans/MCMC analysis while letting $\gamma$ vary freely between $0\leq \gamma\leq 1$ (i.e., the same analysis tested using artificial data).  The classical dSphs all lie well above the Galactic plane and hence are prone to similar levels of diffuse $\gamma$-ray background.  Adopting the background expected for each dSph from the standard Fermi diffuse model of the Milky Way (S. Funk, private communication), we find $\la 7\%$ rms variation from dSph to dSph in estimated background flux.  We neglect these small background variations in the discussion that follows.

For a signal-limited detection (plausible in the case of the Fermi-LAT) $J_{\alpha_c}$ provides an appropriate figure of merit for selecting dSphs as dark matter annihilation targets (filled circles in Fig.~\ref{fig:fig4}).  In that case we find, similarly to \citet{2007PhRvD..75h3526S}, that Draco and Ursa Minor are among the best targets in terms of having the largest median-likelihood estimates for $J_{\alpha_c}$.  Somewhat at odds with that of \citet{2007PhRvD..75h3526S}, our analysis places Sculptor on approximately equal footing with Dra and UMi and perhaps even above UMi given the large error bar associated with that galaxy (of the three, UMi has the smallest available kinematic data set).  This discrepancy likely follows from different choices of prior for $\rho(r)$, as \citet{2007PhRvD..75h3526S} assume NFW profiles.  

On the other hand, in the presence of background, objects with $\alpha_c$ significantly larger than the instrumental point-spread-function are disfavored.  Taking $0.1^{\circ}$ as an optimistic estimate of the angular radius containing $80\%$ of the PSF (close to the optimum integration radius for a point-like source in the background-limited case) of future ground-based $\gamma$-ray detectors in the appropriate energy range (${\cal O}(m_{\chi}^2c^{2}/5)$, \citealt{cta:concept}), we use $J^{\rm  BG}_{\alpha_{c}}\approx J_{\alpha_c}/\sqrt{1+(\alpha_c/0.1^{\circ})^2}$ as a figure of merit in the background-limited case\footnote{We obtain this figure of merit by convolving the dark matter signal with the PSF while assuming that both are Gaussian.  Charbonnier et al. (in prep.) study the effects of angular resolution and demonstrate that within the $80\%$ containment radius, this approximation is accurate to within a few percent.} (empty circles in Fig.~\ref{fig:fig4}). A dominant background significantly changes the ranking of the dSphs, with Leo II now among the best targets. Note, however, that Leo II has the largest error bar as it also has the smallest available kinematic data set of the eight galaxies considered here.

The $J$-factors estimated here do not reach the values required for dark matter detection with existing and near-future instruments in the most conventional particle physics scenarios.  Moreover, most are smaller than the values predicted for the Galactic Center (GC).  To illustrate the last point, we adopt an Einasto profile (e.g., \citealt{navarro04}) for the Milky Way's smooth dark matter component.  We then estimate $J_{\rm GC}\sim 2\times 10^{11}\msun ^2\kpc^{-5}$ toward the GC for $\alpha_{\rm int}=0.01^{\circ}$, and $J_{\rm GC}\sim 1\times 10^{13}\msun^2\kpc^{-5}$ for $\alpha_{\rm int}=0.1^{\circ}$.  However, the presence of strong astrophysical backgrounds on scales of arcminutes \citep{aharonian04}, tens of arcminutes \citep{aharonian06} and tens of degrees \citep{su10} will obfuscate any genuine dark matter signal coming from the direction of the GC.  In contrast, the old, gas-free stellar populations of dSphs provide few sources of point-like or diffuse $\gamma$-ray emission.  On these grounds dSphs provide a favorable alternative to the GC as targets for indirect dark matter searches using instruments with resolution $\alpha_{\rm int}\la 0.1^{\circ}$.  

\begin{table*}
  \caption{Classical dSphs sorted according to their distance \citep{mateo98}. The other columns correspond to twice the half-light radius, the optimum integration angle, the median $\log_{10}(J)$ with 68\% (95\%) confidence intervals for that angle, and background-corrected $\log_{10}(J^{\rm BG})$ (see text).}
\label{tab:tab1}
\begin{center}
\begin{tabular}{@{}lcccccccccc} \hline\hline
  dSph      &~~&   $d$     & $2r_h$ & $\alpha_{c}\approx\frac{2r_h}{d}$&~~& $\log_{10}[J^{\rm med}(\alpha_c)]$ &~~& $\log_{10}[J^{\rm BG}(\alpha_c)]$\\
            &~~&  [kpc]  &  [kpc] &  [deg]      &~~& $[M_{\odot}^{2}\,$kpc$^{-5}]$ &~~& $[M_{\odot}^{2}\,$kpc$^{-5}]$\vspace{0.05cm}\\
\hline
Ursa Minor  &~~&   66    &  0.56  &   0.49      &~~&   $11.9_{-0.1(-0.2)}^{+0.1(+0.6)}$ &~~&  11.2  \vspace{0.1cm}\\
Sculptor    &~~&   79    &  0.52  &   0.38      &~~&   $11.7_{-0.1(-0.1)}^{+0.0(+0.1)}$ &~~&  11.1  \vspace{0.1cm}\\
Draco       &~~&   82    &  0.40  &   0.28      &~~&   $11.8_{-0.1(-0.2)}^{+0.1(+0.2)}$ &~~&  11.3  \vspace{0.1cm}\\
Sextans     &~~&   86    &  1.36  &   0.91      &~~&   $11.1_{-0.4(-0.6)}^{+0.6(+1.1)}$ &~~&  10.1  \vspace{0.1cm}\\
Carina      &~~&   101   &  0.48  &   0.27      &~~&   $10.8_{-0.1(-0.2)}^{+0.1(+0.2)}$ &~~&  10.3  \vspace{0.1cm}\\
Fornax      &~~&   138   &  1.34  &   0.56      &~~&   $11.1_{-0.0(-0.1)}^{+0.1(+0.3)}$ &~~&  10.3  \vspace{0.1cm}\\
LeoII       &~~&   205   &  0.30  &   0.08      &~~&   $11.3_{-0.5(-0.8)}^{+0.8(+1.9)}$ &~~&  11.2  \vspace{0.1cm}\\
LeoI        &~~&   250   &  0.50  &   0.11      &~~&   $10.9_{-0.2(-0.3)}^{+0.1(+0.5)}$ &~~&  10.7  \vspace{0.05cm}\\ 
\hline
\end{tabular}
\end{center}
\end{table*}


\section{Conclusion}

We have shown that our Jeans/MCMC analysis, in which we have allowed the inner slope $\gamma$ to remain a free parameter with possible values between $0\leq \gamma \leq 1$, estimates the astrophysical $J$-factor to within a systematic uncertainty of a factor of $\la 3$ for the Milky Way's classical dSphs, so long as 1) the integration angle is chosen to be $\alpha_{\rm int}=\alpha_{\rm c}\sim 2r_h/d$ and 2) the actual dark matter halo has $\gamma \leq 1$.  For cuspier ($\gamma> 1$) profiles, Figures \ref{fig:fig1} and \ref{fig:fig3} indicate that our analysis under-estimates the $J$-factor systematically (e.g., by a factor of $\ga 10$ if the input halo has $\gamma\ga 1.5$), but such steeply cusped profiles are neither supported by observations nor motivated by current cosmological simulations.  For the family of Einasto profiles applied to dSph-like dark matter halos by \citet{navarro04}, our method over-estimates $J$ by a factor of $\la 2$ (Figure \ref{fig:fig1}).  We conclude that our analysis provides estimates of the $J$-factor that are reliable for the dark matter halo models \citep[e.g.,][and references therein]{2006AJ....132.2685M} most relevant to dSphs.

We consider Figure~\ref{fig:fig4} to provide a good starting point for ranking the classical dSphs as targets for indirect detection of dark matter via annihilation. For a given particle physics scenario the highest median-likelihood values of $J_{\alpha_{c}}$ correspond to the largest expected fluxes.  In the background-limited case, appropriate for searches with future ground-based $\gamma$-ray observatories such as CTA, the highest values of $J^{\rm BG}_{\alpha_{c}}$ correspond to the largest expected statistical significance. Figure~\ref{fig:fig4} illustrates the large differences in the fractional uncertainty in $J_{\alpha_{c}}$ between different dSphs. This uncertainty also forms a criterion for target-selection.  In the end, those dSphs with simultaneously the largest and most tightly constrained $J$-factors have the greatest potential to provide useful constraints on the particle physics of dark matter.

\acknowledgments We thank the referee, Gary Mamon, for providing suggestions that improved the quality of this paper.  We are grateful to Walter Dehnen for providing his code for use in generating artificial dSph data sets.  MGW is supported by NASA through Hubble Fellowship grant HST-HF-51283, awarded by the Space Telescope Science Institute, which is operated by the Association of Universities for Research in Astronomy, Inc., for NASA, under contract NAS 5-26555.  CC acknowledges support from an STFC rolling grant at the University of Leicester.  JAH acknowledges the support of an STFC Advanced Fellowship.  MIW acknowledges the Royal Society for support via a University Research Fellowship. 


\begin{thebibliography}{50}
\expandafter\ifx\csname natexlab\endcsname\relax\def\natexlab#1{#1}\fi

\bibitem[{{Aaronson}(1983)}]{aaronson83}
{Aaronson}, M. 1983, \apjl, 266, L11

\bibitem[{{Abdo et al.}(2010)}]{abdo10}
{Abdo et al.} 2010, \apj, 712, 147

\bibitem[{{Acciari et al.}(2010)}]{acciari10}
{Acciari et al.} 2010, \apj, 720, 1174

\bibitem[{{Aharonian et al.}(2004)}]{aharonian04}
{Aharonian et al.} 2004, \aap, 425, L13

\bibitem[{{Aharonian et al.}(2006)}]{aharonian06}
---. 2006, \nat, 439, 695

\bibitem[{{Aharonian et al.}(2010)}]{aharonian10}
---. 2010, ArXiv:1012.5602

\bibitem[{{Amorisco} \& {Evans}(2011)}]{amorisco10}
{Amorisco}, N.~C., \& {Evans}, N.~W. 2011, \mnras, 411, 2118

\bibitem[{{Atwood} {et~al.}(2009){Atwood}, {Abdo}, {Ackermann}, {Althouse},
  {Anderson}, {Axelsson}, {Baldini}, {Ballet}, {Band}, {Barbiellini}, \&
  et~al.}]{atwood09}
{Atwood}, W.~B., {et~al.} 2009, \apj, 697, 1071

\bibitem[{{Berezinsky} {et~al.}(1992){Berezinsky}, {Gurevich}, \&
  {Zybin}}]{1992PhLB..294..221B}
{Berezinsky}, V.~S., {Gurevich}, A.~V., \& {Zybin}, K.~P. 1992, Physics Letters
  B, 294, 221

\bibitem[{{Binney} \& {Tremaine}(2008)}]{bt08}
{Binney}, J., \& {Tremaine}, S. 2008, {Galactic Dynamics: Second Edition}
  (Princeton University Press)

\bibitem[{{Boyarsky} {et~al.}(2006){Boyarsky}, {Neronov}, {Ruchayskiy},
  {Shaposhnikov}, \& {Tkachev}}]{boyarsky06}
{Boyarsky}, A., {Neronov}, A., {Ruchayskiy}, O., {Shaposhnikov}, M., \&
  {Tkachev}, I. 2006, Physical Review Letters, 97, 261302

\bibitem[{{Boyarsky} {et~al.}(2010){Boyarsky}, {Ruchayskiy}, {Iakubovskyi},
  {Walker}, {Riemer-S{\o}rensen}, \& {Hansen}}]{boyarsky10}
{Boyarsky}, A., {Ruchayskiy}, O., {Iakubovskyi}, D., {Walker}, M.~G.,
  {Riemer-S{\o}rensen}, S., \& {Hansen}, S.~H. 2010, \mnras, 407, 1188

\bibitem[{{Bringmann} {et~al.}(2009){Bringmann}, {Doro}, \&
  {Fornasa}}]{2009JCAP...01..016B}
{Bringmann}, T., {Doro}, M., \& {Fornasa}, M. 2009, Journal of Cosmology and
  Astro-Particle Physics, 1, 16

\bibitem[{{CTA Consortium}(2010)}]{cta:concept}
{CTA Consortium}. 2010, ArXiv:1008.3703

\bibitem[{{Cuddeford}(1991)}]{cuddeford91}
{Cuddeford}, P. 1991, \mnras, 253, 414

\bibitem[{{Einasto} \& {Haud}(1989)}]{einasto89}
{Einasto}, J., \& {Haud}, U. 1989, \aap, 223, 89

\bibitem[{Evans {et~al.}(2004)Evans, Ferrer, \& Sarkar}]{Evans:2003sc}
Evans, N.~W., Ferrer, F., \& Sarkar, S. 2004, Phys. Rev., D69, 123501

\bibitem[{{Gilmore} {et~al.}(2007){Gilmore}, {Wilkinson}, {Wyse}, {Kleyna},
  {Koch}, {Evans}, \& {Grebel}}]{gilmore07}
{Gilmore}, G., {Wilkinson}, M.~I., {Wyse}, R.~F.~G., {Kleyna}, J.~T., {Koch},
  A., {Evans}, N.~W., \& {Grebel}, E.~K. 2007, \apj, 663, 948

\bibitem[{{Goerdt} {et~al.}(2006){Goerdt}, {Moore}, {Read}, {Stadel}, \&
  {Zemp}}]{goerdt06}
{Goerdt}, T., {Moore}, B., {Read}, J.~I., {Stadel}, J., \& {Zemp}, M. 2006,
  \mnras, 368, 1073

\bibitem[{{Hernquist}(1990)}]{hernquist90}
{Hernquist}, L. 1990, \apj, 356, 359

\bibitem[{{Irwin} \& {Hatzidimitriou}(1995)}]{ih95}
{Irwin}, M., \& {Hatzidimitriou}, D. 1995, \mnras, 277, 1354

\bibitem[{{Kleyna} {et~al.}(2002){Kleyna}, {Wilkinson}, {Evans}, {Gilmore}, \&
  {Frayn}}]{kleyna02}
{Kleyna}, J., {Wilkinson}, M.~I., {Evans}, N.~W., {Gilmore}, G., \& {Frayn}, C.
  2002, \mnras, 330, 792

\bibitem[{{Kleyna} {et~al.}(2003){Kleyna}, {Wilkinson}, {Gilmore}, \&
  {Evans}}]{kleyna03}
{Kleyna}, J.~T., {Wilkinson}, M.~I., {Gilmore}, G., \& {Evans}, N.~W. 2003,
  \apjl, 588, L21

\bibitem[{{Koch} {et~al.}(2007){Koch}, {Wilkinson}, {Kleyna}, {Gilmore},
  {Grebel}, {Mackey}, {Evans}, \& {Wyse}}]{koch07}
{Koch}, A., {Wilkinson}, M.~I., {Kleyna}, J.~T., {Gilmore}, G.~F., {Grebel},
  E.~K., {Mackey}, A.~D., {Evans}, N.~W., \& {Wyse}, R.~F.~G. 2007, \apj, 657,
  241

\bibitem[{{Kuhlen}(2010)}]{kuhlen10}
{Kuhlen}, M. 2010, Advances in Astronomy, 2010

\bibitem[{{Kuhlen} {et~al.}(2008){Kuhlen}, {Diemand}, \& {Madau}}]{kuhlen08}
{Kuhlen}, M., {Diemand}, J., \& {Madau}, P. 2008, \apj, 686, 262

\bibitem[{{Lake}(1990)}]{1990Natur.346...39L}
{Lake}, G. 1990, \nat, 346, 39

\bibitem[{{{\L}okas} {et~al.}(2005){{\L}okas}, {Mamon}, \& {Prada}}]{lokas05}
{{\L}okas}, E.~L., {Mamon}, G.~A., \& {Prada}, F. 2005, \mnras, 363, 918

\bibitem[{{Martinez} {et~al.}(2009){Martinez}, {Bullock}, {Kaplinghat},
  {Strigari}, \& {Trotta}}]{martinez09}
{Martinez}, G.~D., {Bullock}, J.~S., {Kaplinghat}, M., {Strigari}, L.~E., \&
  {Trotta}, R. 2009, JCAP, 6, 14

\bibitem[{{Mateo} {et~al.}(2008){Mateo}, {Olszewski}, \& {Walker}}]{mateo08}
{Mateo}, M., {Olszewski}, E.~W., \& {Walker}, M.~G. 2008, \apj, 675, 201

\bibitem[{{Mateo}(1998)}]{mateo98}
{Mateo}, M.~L. 1998, \araa, 36, 435

\bibitem[{{Merritt} {et~al.}(2006){Merritt}, {Graham}, {Moore}, {Diemand}, \&
  {Terzi{\'c}}}]{2006AJ....132.2685M}
{Merritt}, D., {Graham}, A.~W., {Moore}, B., {Diemand}, J., \& {Terzi{\'c}}, B.
  2006, \aj, 132, 2685

\bibitem[{{Mu{\~n}oz et al.}(2006)}]{munoz06}
{Mu{\~n}oz et al.} 2006, \apj, 649, 201

\bibitem[{{Navarro} {et~al.}(2004){Navarro}, {Hayashi}, {Power}, {Jenkins},
  {Frenk}, {White}, {Springel}, {Stadel}, \& {Quinn}}]{navarro04}
{Navarro}, J.~F., {et~al.} 2004, \mnras, 349, 1039

\bibitem[{{Navarro, Frenk \& White}(1996)}]{navarro96}
{Navarro, Frenk \& White}. 1996, \apj, 462, 563

\bibitem[{{Navarro, Frenk \& White}(1997)}]{navarro97}
---. 1997, \apj, 490, 493

\bibitem[{Pieri {et~al.}(2009)}]{pieri09}
Pieri, L., {et~al.} 2009, \aap, 496, 351

\bibitem[{{Sohn} {et~al.}(2007){Sohn}, {Majewski}, {Mu{\~n}oz}, {Kunkel},
  {Johnston}, {Ostheimer}, {Guhathakurta}, {Patterson}, {Siegel}, \&
  {Cooper}}]{sohn07}
{Sohn}, S.~T., {et~al.} 2007, \apj, 663, 960

\bibitem[{{Springel} {et~al.}(2008){Springel}, {Wang}, {Vogelsberger},
  {Ludlow}, {Jenkins}, {Helmi}, {Navarro}, {Frenk}, \& {White}}]{springel08}
{Springel}, V., {et~al.} 2008, \mnras, 391, 1685

\bibitem[{{Stoehr} {et~al.}(2003){Stoehr}, {White}, {Springel}, {Tormen}, \&
  {Yoshida}}]{stoehr03}
{Stoehr}, F., {White}, S.~D.~M., {Springel}, V., {Tormen}, G., \& {Yoshida}, N.
  2003, \mnras, 345, 1313

\bibitem[{{Strigari} {et~al.}(2008{\natexlab{a}}){Strigari}, {Bullock},
  {Kaplinghat}, {Simon}, {Geha}, {Willman}, \& {Walker}}]{2008Natur.454.1096S}
{Strigari}, L.~E., {Bullock}, J.~S., {Kaplinghat}, M., {Simon}, J.~D., {Geha},
  M., {Willman}, B., \& {Walker}, M.~G. 2008{\natexlab{a}}, \nat, 454, 1096

\bibitem[{{Strigari} {et~al.}(2007){Strigari}, {Koushiappas}, {Bullock}, \&
  {Kaplinghat}}]{2007PhRvD..75h3526S}
{Strigari}, L.~E., {Koushiappas}, S.~M., {Bullock}, J.~S., \& {Kaplinghat}, M.
  2007, \prd, 75, 083526

\bibitem[{{Strigari} {et~al.}(2008{\natexlab{b}}){Strigari}, {Koushiappas},
  {Bullock}, {Kaplinghat}, {Simon}, {Geha}, \& {Willman}}]{strigari08b}
{Strigari}, L.~E., {Koushiappas}, S.~M., {Bullock}, J.~S., {Kaplinghat}, M.,
  {Simon}, J.~D., {Geha}, M., \& {Willman}, B. 2008{\natexlab{b}}, \apj, 678,
  614

\bibitem[{{Su} {et~al.}(2010){Su}, {Slatyer}, \& {Finkbeiner}}]{su10}
{Su}, M., {Slatyer}, T.~R., \& {Finkbeiner}, D.~P. 2010, \apj, 724, 1044

\bibitem[{{The MAGIC Collaboration}(2011)}]{seg1magic11}
{The MAGIC Collaboration}. 2011, ArXiv:1103.0477

\bibitem[{{Walker} {et~al.}(2007){Walker}, {Mateo}, {Olszewski}, {Gnedin},
  {Wang}, {Sen}, \& {Woodroofe}}]{walker07b}
{Walker}, M.~G., {Mateo}, M., {Olszewski}, E.~W., {Gnedin}, O.~Y., {Wang}, X.,
  {Sen}, B., \& {Woodroofe}, M. 2007, \apjl, 667, L53

\bibitem[{{Walker} {et~al.}(2009){Walker}, {Mateo}, {Olszewski},
  {Pe{\~n}arrubia}, {Evans}, \& {Gilmore}}]{walker09d}
{Walker}, M.~G., {Mateo}, M., {Olszewski}, E.~W., {Pe{\~n}arrubia}, J.,
  {Evans}, N.~W., \& {Gilmore}, G. 2009, \apj, 704, 1274

\bibitem[{{Walker, Mateo \& Olszewski}(2009)}]{walker09a}
{Walker, Mateo \& Olszewski}. 2009, \aj, 137, 3100

\bibitem[{{Wolf} {et~al.}(2010){Wolf}, {Martinez}, {Bullock}, {Kaplinghat},
  {Geha}, {Mu{\~n}oz}, {Simon}, \& {Avedo}}]{wolf10}
{Wolf}, J., {Martinez}, G.~D., {Bullock}, J.~S., {Kaplinghat}, M., {Geha}, M.,
  {Mu{\~n}oz}, R.~R., {Simon}, J.~D., \& {Avedo}, F.~F. 2010, \mnras, 406, 1220

\bibitem[{{Zhao}(1996)}]{zhao96}
{Zhao}, H. 1996, \mnras, 278, 488

\end{thebibliography}

\end{document}